\documentclass[preprint,12pt]{elsarticle}

\usepackage{amssymb}
\usepackage{amsmath}
\usepackage{booktabs}    
\usepackage{longtable}
\usepackage{listings}
\usepackage{multirow}
\usepackage{hyperref} 

\journal{Applied Soft Computing}

\begin{document}

\begin{frontmatter}



\title{Generalized Distribution Estimation for Asset Returns}


\author[inst1]{Ísak Pétursson\corref{mycorrespondingauthor}}
\cortext[mycorrespondingauthor]{Corresponding author}\ead{isakp23@ru.is}%% Author name
\author[inst1,inst2]{María Óskarsdóttir}
\affiliation[inst1]{organization={Reykjavik University, Department of Computer Science},%Department and Organization
            addressline={Menntavegur 1}, 
            city={Reykjavik},
            postcode={102}, 
            state={Captial Region},
            country={Iceland}}
\affiliation[inst2]{organization={University of Southampton, School of Mathematical Sciences},%Department and Organization
            addressline={University Road}, 
            city={Southampton},
            postcode={SO17 1BJ}, 
            state={Hampshire},
            country={United Kingdom}}
%% Abstract
\begin{abstract}
Accurately modeling the distribution of asset returns enables better risk management, portfolio optimization, and financial decision-making especially in uncertain and volatile markets. This paper presents a novel approach to estimating the distribution of asset log returns using quantile regression combined with smoothed density estimation. Our proposed model, qLSTM, leverages asset-neutral features to deliver robust and generalizable predictions across diverse asset classes, outperforming baseline and dense models in quantile loss metrics for real-world data. A hybrid model, averaging Gaussian and model-predicted quantiles, qHybrid, further improves performance on synthetic datasets with skewed and heavy-tailed distributions. Evaluation using metrics like Wasserstein distance, CRPS, and VaR demonstrates the models' effectiveness in capturing tail risks and broader distributional characteristics.
\end{abstract}

%\includegraphics[width=\textwidth]{ASOC_graphicalabstract.png}

%\item Quantile model generalizes return distribution estimation for varying time sequences.
%\item Hybrid model integrates Gaussian and ML-predicted quantiles for heavy-tailed data.
%\item Metrics (Wasserstein, CRPS, VaR) validate tail risk and distributional modeling.
%\item Model performs robustly across real-world and synthetic financial datasets.
%\item Public datasets and code promote reproducibility and future research.

%% Keywords

\end{frontmatter}


\section{Introduction}\label{sec1}
In the financial world, accurate modeling of asset price movements and their associated risks is a cornerstone of decision-making for investors, portfolio managers, and risk analysts \cite{fama-capm}. This modeling forms the foundation of derivative pricing, risk management, and portfolio optimization. One widely used approach assumes that asset price changes over time can be modeled as a stochastic process, with the distribution of returns playing a central role in these models \cite{bs-prime}.

Machine learning has transformed several practices and processes withing the finance domain by enabling models that capture the complex, nonlinear relationships inherent in market data \cite{ml-finance-review}. Applications range from price and return prediction to risk management and portfolio optimization, leveraging methods such as ARIMA, LSTM, and CNNs \citep{arima1, lstm1, cnn1}. However, much of this research focuses on point predictions rather than estimating the full return distribution, which is essential for understanding uncertainty and tail risks. 

Traditionally, logarithmic asset returns are often assumed to follow a normal distribution; a famous example is the Black-Scholes model, which is used to find the fair value of option derivatives \cite{bs-prime}. However, in practice, this assumption frequently does not hold \cite{non-normal}. Returns often display non-normal characteristics, such as skewness or heavy tails. Consequently, models that rely on normality assumptions can lead to suboptimal or even incorrect results \cite{non-normal-bad}. To address this issue, some approaches generalize assumptions or adopt alternative distributions, such as the Student's $t$- distribution, which accommodates skewness, kurtosis, and heavy tails \cite{student-t}. Furthermore, distributional approaches in machine learning, show recent advances in conditional modeling and quantile regression, address this gap by providing a stronger method for distribution estimation \citep{non-cond-dist-2, cond-dist-1}. 

More recently, a new approach has emerged, namely to model the distribution directly \cite{dist-pred}. This approach offers several key advantages. A more accurate model of the return distribution could improve derivative pricing, inform risk management, and enhance portfolio optimization. Although the proposed method shows promising results, it only considers one day's worth of data and has a fixed look ahead period of 22 days. Furthermore, the method relies heavily on  derived features which are exclusive to stocks, meaning the method does not generalize to other asset classes.

In this paper, we propose a model that estimates the distribution of returns given current observations. Our model predicts the quantiles of the log return distribution over $n$ days given $n$ days worth of data, thus extending previous work\cite{dist-pred}. Although quantile regression is not a novel concept, we convert these quantiles into a full distribution using a smoothed density estimation function. This enables a more precise estimation of the return distribution over a given time period. The model is trained on a dataset of asset returns and general market variables, ensuring that each characteristic is asset-class neutral, which makes the approach applicable across different asset classes.

This paper contributes to the literature in the following ways:
\begin{itemize}
    \item Proposes a generalizable distribution estimation model that:
    \begin{itemize}
        \item Uses only asset-neutral features.
        \item Allows the consideration of multiple periods of return.
    \end{itemize}
    \item Proposes a hybrid model that utilizes standard Gaussian quantiles in combination with the generalized model.
    \item Introduces an experimental framework for enhanced reproducibility in the field by using only data which is is publicly available.
\end{itemize}
The results are compared in two ways: quantile loss is used to evaluate the performance of our approaches against a baseline linear quantile regression model and the original dense approach in \cite{dist-pred}. We find that our proposed qLSTM model achieves the lowest overall quantile loss for real-world data, demonstrating its robustness and versatility across diverse asset classes, including cryptocurrencies, commodities, and equity indices. The hybrid model, qHybrid, performs particularly well on synthetic data, especially for distributions with heavy tails or skewness, highlighting its flexibility. Additionally, we evaluate the models using standard metrics such as the Wasserstein distance and the Continuous Ranked Probability Score (CRPS). These metrics reveal that qLSTM excels in capturing extreme tail risks, while the qHybrid model achieves better calibration for broader distributions.

The rest of this paper is structured as follows. In Section \ref{sec2}, we provide an overview of the existing literature on machine learning in finance, focusing on the state of machine learning in finance, distributional modeling, and quantile regression methods. Section \ref{sec3} outlines our methodology, including the problem formulation, model architecture, and training procedure. The data sources, including real-world and synthetic datasets, are described in detail, along with feature engineering and hyperparameter optimization. Section \ref{sec4} presents the results of our experiments, comparing the performance of our proposed models with existing baselines using various metrics. Finally, Section \ref{sec5} concludes the paper by summarizing our contributions and findings.

\section{Literature Review}\label{sec2}
The application of machine learning to estimate phenomena within finance is well-explored, with much of the literature focusing on optimizing returns by predicting directions, prices, or returns. As noted by Dessain \cite{loss-meta}, the exponential growth in academic studies has made it increasingly difficult to synthesize existing knowledge within the domain. Similarly, Strader \textit{et al}. \cite{meta_on_ML} and Shah \textit{ et al.}. \cite{w-meta} emphasize the diversity of methods, such as ARIMA \citep{arima1, arima2}, LSTM \citep{lstm1, lstm2}, CNN \citep{cnn1, cnn2}, and hybrid models \citep{hybrid1, hybrid2}, as well as the inconsistency in the benchmarks, which complicates the comparisons between studies. To address these issues, our study uses publicly available data to ensure reproducibility and comparability.

In the domain of distribution estimation, significant advances have been made in non-conditional density estimation \citep{non-cond-dist-1, non-cond-dist-2, non-cond-dist-3}. In contrast, this work focuses on conditional distributions allowing us to take various features from both the asset in question as well as features from the market to inform the shape of the distribution. Hu and Nan propose a method for estimating conditional distributions by modeling the hazard function and deriving the cumulative distribution function (CDF) via integration \cite{cond-dist-1}, while Feindt \cite{cond-dist-3} demonstrates the utility of CDF-based methods across domains, including finance. However, both approaches prioritize global density estimation. Our work differs by directly estimating the quantile function, emphasizing financial applications, and incorporating domain-specific knowledge to capture tail risks.

Another notable contribution is by Rasul \textit{ et al.}. \cite{non-cond-dist-2}, who use normalizing flows conditioned on time series data for multistep distributional estimation, where the conditional distribution of \( r_{k} \) depends not only on past observations \( X \), but also on intermediate estimations \( r_{1}, \ldots, r_{k-1} \). This allows uncertainty to propagate across horizons. In contrast, we assume that each return on the forecast trajectory is drawn from the same distribution (i.e., \( r_{1} \sim f(X), r_{2} \sim f(X)\)), simplifying the computation while focusing on quantile estimation and tail risk for financial applications.

Focusing specifically on finance, Hickling and Prangle \citep{tails} leverage extreme value theory inspired transformations within normalizing flows to model heavy-tailed distributions. Similarly, Barun and Liu. \cite{sub-dist-Liue} utilize quantile regression to address distributional challenges in financial returns. Building on this foundation, our work incorporates long-short-term memory (LSTM) networks to estimate quantiles over \( n \) day horizons, ensuring robustness across asset classes by utilizing asset-neutral features, in contrast to feature-heavy approaches like those in \cite{dist-pred}, which rely on stock-specific characteristics and macroeconomic variables.

Despite progress, challenges remain in balancing model complexity, interpretability, and applicability across asset classes. This work addresses gaps in tail risk modeling by combining quantile-based LSTM architectures with asset-neutral features, providing robust and generalizable estimations across diverse financial datasets.

\section{Methodology}\label{sec3}
Our proposed methodology consists of several steps. First we formalize the problem which this paper tackles, then we describe our two stage model to take into account both asset specific features as well as market general features. It is then followed by details about the two stage quantile loss function deployed. Next, we describe both real and synthetic data sources used in the training and evaluation of the models. Last, we detail the experimental setup and the metrics used for evaluation. 

The code used to implement all the approaches in this section, as well as to evaluate the models, is publicly available at \href{https://github.com/izzak98/dist-pred}{https://github.com/izzak98/dist-pred}

\subsection{Problem Definition}

An asset is represented as a matrix of input features \( X \in \mathbb{R}^{m \times t} \), where \( m \) is the number of features, and \( t \) is the number of historical time steps. The matrix \( X \) is structured as follows
\begin{equation*}
X = \begin{bmatrix}
x_{1,1} & x_{1,2} & x_{1,3} & \cdots & x_{1,t} \\
x_{2,1} & x_{2,2} & x_{2,3} & \cdots & x_{2,t} \\
x_{3,1} & x_{3,2} & x_{3,3} & \cdots & x_{3,t} \\
\vdots & \vdots & \vdots & \ddots & \vdots \\
x_{m,1} & x_{m,2} & x_{m,3} & \cdots & x_{m,t} \\
\end{bmatrix}
\end{equation*} 
where $x_{i,j}$ is the value of feature $i$ at time $j$.
Accompanying the asset matrix is the market data for the same \( t \) time steps, represented as a matrix \( Z \in \mathbb{R}^{g \times t} \), where \( g \) is the number of market features, that is
\begin{equation*}
Z = \begin{bmatrix}
z_{1,1} & z_{1,2} & \cdots & z_{1,t} \\
z_{2,1} & z_{2,2} & \cdots & z_{2,t} \\
\vdots & \vdots & \ddots & \vdots \\
z_{g,1} & z_{g,2} & \cdots & z_{g,t} \\
\end{bmatrix}
\end{equation*}
where $z_{i,j}$ is value of market asset $i$ at time $j$.

Our goal is to estimate the distribution of the log returns of the asset for the future time steps from \( 1 \) to \( t \), where \( t \) represents both the number of time steps used for the estimation and the forecast horizon. The vector of log returns \( \vec{r} \in \mathbb{R}^t \) is defined as:
\begin{equation*}
\vec{r} = \begin{bmatrix}
r_{1} \\
r_{2} \\
\vdots \\
r_{n} \\
\end{bmatrix}.
\end{equation*}

Given the input features \( X \) and \( Z \), the objective is to model and estimate the distribution of log returns, that is,
\begin{equation*}
\vec{r} \sim f(X, Z)
\end{equation*}
where \( f(X, Z) \) represents the model for the estimation of log returns based on the asset and market features.

\subsection{Model Architecture}
To model the distribution of log returns, our objective is to estimate the quantiles \( \vec{\tau} \) of the log return distribution over the estimation horizon. Utilizing density estimation, these can later be converted into distributions. The quantile vector is defined as
\begin{equation} \label{eq:taus}
\vec{\tau} = \begin{bmatrix}
\tau_1 \\
\tau_2 \\
\vdots \\
\tau_{\mathcal{T}} \\
\end{bmatrix}
\end{equation}
where \( \mathcal{T} \) denotes the number of quantiles. For the purpose of this research, we chose quantiles that represent broad ranging characteristics of a distribution, including quartiles and deciles as well as more fine grained intervals at the lower and upper end of the distribution to capture heavy tails. The quantiles used in this research can be found in \ref{app:qunats}.

For this task, we use long- and short-term memory (LSTM) networks, a type of recurrent neural network (RNN) \cite{lstm-prime}, which are well suited for sequential data modeling due to their ability to capture long-term dependencies and are less prone to issues such as vanishing gradients. LSTMs maintain two states at each time step: the cell state \( \vec{c}_t \) and the hidden state \( \vec{h}_t \), which are updated using gating mechanisms. In addition there are the input and output gates, \( \vec{i}_t \) and \( \vec{o}_t \), respectively. Specifically, the updates are governed by the equations:
\begin{equation*}
\begin{aligned}
    \vec{i}_t &= \sigma(W_i \vec{x}_t + U_i h_{t-1} + \vec{b}_i), \\
    \vec{f}_t &= \sigma(W_f \vec{x}_t + U_f \vec{h}_{t-1} + \vec{b}_f), \\
    \vec{o}_t &= \sigma(W_o \vec{x}_t + U_o \vec{h}_{t-1} + \vec{b}_o), \\
    \tilde{\vec{c}}_t &= \tanh(W_c \vec{x}_t + U_c \vec{h}_{t-1} + \vec{b}_c), \\
    \vec{c}_t &= \vec{f}_t \odot \vec{c}_{t-1} + \vec{i}_t \odot \tilde{\vec{c}}_t, \\
    \vec{h}_t &= \vec{o}_t \odot \tanh(\vec{c}_t),
\end{aligned}
\end{equation*}
where \( \sigma(\cdot) \) and \( \tanh(\cdot) \) are the activation functions of the sigmoid and hyperbolic tangent, respectively, and \( \odot \) denotes the multiplication of elements. Unlike many other time-series models, LSTMs can flexibly handle variable-length sequences, making them appropriate for our task as we are working with a variable number of sequence lengths.

Loosely following the methodology outlined in \cite{dist-pred}, we adopt a two-stage approach to estimate the distribution of future log returns.

\subsubsection{Stage 1: Asset-Specific Quantile Estimation}
In the first stage, the model focuses on estimating the quantiles of future normalized returns $\vec{\tilde{r}}$ by leveraging the asset-specific features \( X \). This is done by estimating the quantiles \( Q_{\vec{\tilde{r}}}^\tau \) of the volatility-normalized returns \( \vec{\tilde{r}} \), where
\begin{equation*}
\vec{\tilde{r}} = \frac{\vec{r}}{\vec{\bar{\sigma}}}
\end{equation*}
with 
\begin{equation*}
\vec{r} = \begin{bmatrix}
r_{1} \\
\vdots \\
r_{t}
\end{bmatrix} \quad \textrm{and} \quad 
\vec{\bar{\sigma}} = \begin{bmatrix}
\bar{\sigma}_{1} \\
\vdots \\
\bar{\sigma}_{t}
\end{bmatrix}
\end{equation*}
Here, \( \vec{\bar{\sigma}} \) represents the average volatility of assets in the same group at some time step, where \( t \) is the estimation horizon. 

The average volatility \( \bar{\sigma}_t \) is calculated as the exponentially weighted moving average (EWMA) of the log returns across all \( K \) assets in the group, with a decay factor of 0.94, in line with standard practices and the methodology from \cite{dist-pred}. Formally the expression is
\begin{equation*}
\sigma_t^2 = \lambda \sigma_{t-1}^2 + (1 - \lambda) r_{t-1}^2
\end{equation*}
where $\lambda=0.94$ is the decay factor. 
Here, $\bar{\sigma}_t$ is calculated as 
\begin{equation*}
\bar{\sigma}_t = \frac{1}{K} \sum_{i=1}^{K} \sigma_{i,t}
\end{equation*}
where \( \sigma_{i,t} \) is the volatility of asset \( i \) at time \( t \).
We represent this first stage with
\begin{equation*}
f_1(X) =  Q_{\vec{\tilde{r}}}^\tau , 
\end{equation*}
that is, a function of the input features gives the quantiles of the volatility-normalized log returns.

\subsubsection{Stage 2: Market Data Scaling}
In the second stage, the model incorporates the broader market data features \( Z \) to refine the estimation from the first stage. The market data is used to create a scaling factor \( \vec{\tilde{\sigma}} \), which adjusts the asset-specific quantile estimation based on market conditions.

The scaling factor \( \vec{\hat{\sigma}} \) is generated by applying a transformation to the market feature matrix \( Z \), capturing the influence of market conditions on future returns. This results in the final quantile estimation for the asset's log returns:
\begin{equation}\label{eq:qlstm}
Q_{\vec{r}}^\tau = \vec{\hat{\sigma}} \cdot  Q_{\vec{\tilde{r}}}^\tau
\end{equation}
where \( Q_{\vec{r}}^\tau \) represents the final estimated quantile for the asset's log returns, \( \vec{\bar{\sigma}} \) is the scaling factor derived from the market data features \( Z \), which adjusts the asset-specific estimations based on the broader market environment, and \( Q_{\vec{\tilde{r}}}^\tau \) is the asset-specific quantile estimation from Stage 1, normalized by the asset's volatility.
We can formally represent this stage as
\begin{equation*}
f_2(Z) = \vec{\hat{\sigma}},
\end{equation*}
that is, a function of the market data gives a scalar value to scale the quantiles from \(f_1(X)\).

Now let $f_{\text{DE}}$ be a Quantile-Smooth Density Estimation. This method takes quantiles and their associated probabilities to generate a smooth probability density function (PDF) while ensuring that the CDF is monotonic, bounded, and normalized. It smooths the density to prevent spikes, enforces non-negativity, and normalizes the PDF to integrate to one, providing a well-behaved representation of the underlying distribution; for more details see the appendix \ref{app:DE}. This leads to the final model which is given by the representation
\begin{equation*}
\vec{r} \sim f(X, Z) = f_{\text{DE}}\left[f_1(X) \cdot f_2(Z) \right]
\end{equation*}
where $f_{\text{DE}}$ represents the Quantile-Smooth Density Estimation. 
We refer to this model as qLSTM.

\paragraph{Hybrid model}
We also propose a hybrid model, which we refer to as qHybrid, where after training the qLSTM model, we compute the average of the quantile of the model and a Gaussian fit of the log returns in the lookback period, $-t$ to $0$ used in $X$. Formally, we define this as
\begin{equation} \label{eq:qhybrid}
Q_{\mathcal{H}}^\tau = \frac{Q_{\mathcal{N}}^\tau + Q_{\vec{\tilde{r}}}^\tau}{2},
\end{equation}
where \( Q_{\mathcal{N}}^\tau \) represents the Gaussian quantiles across the $\vec{\tau}$ defined in Eq. \eqref{eq:taus}, computed as
\begin{equation*}
Q_{\mathcal{N}}^\tau = [\mu + \sigma \Phi^{-1}(\tau_i)]_{i=1}^\mathcal{T},
\end{equation*}
with \( \Phi(x) \) being the standard normal cumulative distribution function:
\begin{equation*}
\Phi(x) = \frac{1}{\sqrt{2\pi}} \int_{-\infty}^x \exp\left(-\frac{t^2}{2}\right) \, dt.
\end{equation*}
Here, \( \mu \) and \( \sigma \) denote the mean and standard deviation of the observed returns in the lookback period, from \(-t\) to \(0\).

\subsubsection{Loss Function}

Following the methodology of \cite{dist-pred}, we use a custom loss function based on quantile regression to train the model. The original loss function is designed to minimize the error between the true returns and the estimated quantiles. It is defined as

\begin{equation} \label{eq:loss1}
    \mathcal{L}_{\tau} = \frac{1}{B K} \sum_{\tau \in \mathcal{T}} \sum_{i=1}^B \left[ \rho_{\tau}\left( r_{i} - \hat{Q}_{r_{i}}(\tau) \right) + \rho_{\tau}\left( \tilde{r}_{i} - \hat{Q}_{\tilde{r}_{i}}(\tau) \right) \right] 
\end{equation}
where \( \mathcal{T} \) is the set of quantiles used for the estimation, \( B \) is the batch size (i.e., the number of assets in each batch), \( K \) is the number of quantiles in the set \( \mathcal{T} \), \( r_{i} \) is the raw return of the asset, \( \tilde{r}_{i} \) is the standardized (volatility-normalized) return of the asset and \( \hat{Q}_{r_{i}}(\tau) \) and \( \hat{Q}_{\tilde{r}_{i}}(\tau) \) are the estimated quantiles of the raw and standardized returns for quantile \( \tau \), respectively
Furhtermore, \( \rho_{\tau}(\xi) \) is the quantile loss function
\begin{equation*}
    \rho_{\tau}(\xi) = \begin{cases}
        \tau \cdot \xi, & \text{if } \xi \geq 0 \\
        (\tau - 1) \cdot \xi, & \text{if } \xi < 0
    \end{cases}
\end{equation*}
where \( \xi \) represents the error (residual) between the true return value and the estimated quantile.

\subsubsection{Modified Loss Function}
Our proposed model generalizes the model in \cite{dist-pred} by generating a singe estimation for log returns over a look-ahead window, and we therefore modify the loss function in Eq.\eqref{eq:loss1} to account for each future time step \( t \in \{1, 2, \dots, T\} \) within the estimation horizon. The modified loss function is defined as

\begin{equation*}
    \mathcal{L}_{\mathcal{T}} = \frac{1}{B T K} \sum_{\tau \in \mathcal{T}} \sum_{t=1}^T \sum_{i=1}^B \left[ \rho_{\tau}\left( r_{i,t} - \hat{Q}_{\vec{r}_{i}}(\tau) \right) + \rho_{\tau}\left( \tilde{r}_{i,t} - \hat{Q}_{\vec{\tilde{r}}_{i}}(\tau) \right) \right]
\end{equation*}
where \( T \) is the number of future time steps (i.e., the estimation horizon) in the look-ahead window, \( r_{i,t} \) and \( \tilde{r}_{i,t} \) are the raw and standardized returns of the asset at future time step \( t \), and \( \hat{Q}_{\vec{r}_{i,t}}(\tau) \) and \( \hat{Q}_{\vec{\tilde{r}}_{i,t}}(\tau) \) are the estimated quantiles of the raw and standardized returns for asset \( i \) at future time step \( t \).

This modification allows us to compute the loss over the entire estimation window, training the model to minimize the quantile loss for each future time step within the horizon. The inclusion of both raw and standardized returns helps ensure that the model learns both absolute and volatility-adjusted relationships.
\subsection{Market Data} \label{subsec:data}

We utilize a diverse set of financial assets and market data for our analysis. The assets span various categories and markets, ensuring comprehensive coverage. The data spans from February 3, 2000 up to January 1, 2024, sourced from the Yahoo Finance API. All data are normalized using rolling Z-scores, with the window length set as a hyperparameter to be optimized during model training.

\subsubsection{Asset Selection}
Several asset classes are chosen to create a diverse representation to emphasize the generalizability of the model. An overview can be found in table \ref{tab:asset_selection}. Four asset classes are chosen: stocks, currencies, commodities, and cryptocurrencies to capture a diverse set of asset classes to test the robustness of the model. Within stocks there are three sub-asset classes for the index the stocks are listed on, S\&P 500, Euro Stoxx 50, and Nikkei 225. Full details on the selected assets as well as summary statistics, can be found in Appendix \ref{app:assets}.
\begin{table}[h]
    \centering
    \caption{Summary of selected assets}
    \begin{tabular}{ll}
        \toprule
        {Asset Class} & {Selection} \\ \midrule
        {Stocks (S\&P 500)} & 10 stocks from the S\&P 500 index. \\ 
        {Stocks (Euro Stoxx 50)} & 10 stocks from the Euro Stoxx 50 index. \\ 
        {Stocks (Nikkei 225)} & 10 stocks from the Nikkei 225 index. \\ 
        {Currencies} & 10 currency pairs. \\ 
        {Commodities} & 10 commodities. \\ 
        {Cryptocurrencies} & 10 cryptocurrencies. \\ \bottomrule
    \end{tabular}
    \label{tab:asset_selection}
\end{table}

\subsubsection{Feature Engineering}
Additional features are computed for each asset to enhance the models' ability to capture various market dynamics. Table \ref{tab:feature_engineering} shows an overview of the engineered features. These features span returns, volatility metrics, moving averages, technical indicators, risk metrics, and the asset class to which the asset belongs. These features are common practice in technical analysis and are widely used across studies \citep{ti-1, ti-2, ti-3, ti-4}. 
\begin{table}[h]
    \centering
    \caption{Overview of engineered features}
    \begin{tabular}{lp{9cm}}
        \toprule
        {Feature Category} & {Features} \\ \midrule
        {Returns} & 2-day, 5-day, 22-day returns, cumulative returns, log returns. \\ 
        {Volatility Metrics} & 2-day, 5-day, 22-day volatility, skewness, kurtosis. \\ 
        {Moving Averages} & 2-day, 5-day, 22-day simple and exponential moving averages (SMA, EMA). \\ 
        {Technical Indicators} & RSI, MACD, Bollinger Bands, Stochastic Oscillator, VWAP. \\ 
        {Risk Metrics} & 2-day, 5-day, 22-day Sharpe Ratio. \\ 
        {Categorical Features} & Asset category. \\ \bottomrule
    \end{tabular}
    \label{tab:feature_engineering}
\end{table}

\subsubsection{Market Data}
In addition to the assets, general market data is incorporated to capture broader trends and enhance the model's ability to account for macroeconomic factors. An overview of the market data sources can be found in Table \ref{tab:market_data}. This data includes stock indices, commodities, other indices, and treasury yields. For more details on market data, see Appendix \ref{app:market-data}.

\begin{table}[h]
    \centering
    \caption{Overview of market data sources}
    \begin{tabular}{lp{9cm}}
        \toprule
        {Data Category} & {Data Source} \\ \midrule
        {Stock Indices} & S\&P 500, NASDAQ, Dow Jones, Russell 2000, Euro Stoxx 50, Nikkei 225. \\ 
        {Commodities} & Crude oil, gold, silver futures. \\ 
        {Other Indices} & Baltic Dry Index, VIX, US Dollar Index, Corporate Bond Index (Bloomberg Barclays). \\ 
        {Treasury Yields} & 10-year and 3-month Treasury yields. \\ \bottomrule
    \end{tabular}
    \label{tab:market_data}
\end{table}

\subsection{Synthetic Data}
In addition to evaluating the model on real asset data, described in section \ref{subsec:data}, we also test its performance on synthetic datasets generated from several probability distributions, including normal, gamma, log-normal, and uniform. These synthetic datasets simulate various market conditions, where the returns are sampled from distribution $\mathcal{D}$, and then scaled according to the desired mean \( \ddot{\mu} \) and standard deviation \( \ddot{\sigma} \) as follows:

\begin{equation*}
r_{\text{synthetic}} = \mathcal{D} \cdot \ddot{\sigma} + \ddot{\mu}
\end{equation*}

We generate the synthetic return data using the following distributions:

\paragraph{Normal Distribution}
   \begin{equation*}
   \mathcal{D} \sim \mathcal{N}(0, 1)
   \end{equation*}
   where \( \mathcal{N}(0, 1) \) represents a standard normal distribution with mean 0 and standard deviation 1.

\paragraph{Gamma Distribution}
   \begin{equation*}
   \mathcal{D} \sim \Gamma(2, 1)
   \end{equation*}
   where \( \Gamma(2, 1) \) represents a gamma distribution with shape parameter 2 and scale parameter 1. The shape and scale parameters were selected to provide variability while covering a realistic range of potential returns. The distribution is then standardized to have a mean of 0 and a variance of 1.

\paragraph{Log-normal Distribution}
   \begin{equation*}
   \mathcal{D} \sim \text{Lognormal}(0, 1)
   \end{equation*}
   where the log-normal distribution is parameterized by a mean of 0 and a standard deviation of 1 for the underlying normal distribution, this too is standardized before applying the scaling.

\paragraph{Uniform Distribution}
   \begin{equation*}
   \mathcal{D} \sim \mathcal{U}(-1, 1)
   \end{equation*}
   representing a uniform distribution between -1 and 1.

\paragraph{}
For each distribution, the synthetic returns \( r_{\text{synthetic}} \) are scaled by the randomly sampled mean \( \ddot{\mu} \sim \text{Uniform}(-0.001, 0.001) \) and standard deviation \( \ddot{\sigma} \sim \text{Uniform}(0.01, 0.03) \). This scaling ensures that the synthetic datasets mimic realistic market conditions across a variety of asset types.

We generate 10 synthetic datasets per distribution, each consisting of 1000 samples, to assess the model's robustness under these simulated conditions. Additionally, we generate \( g \) synthetic datasets for the market data, where \( g \) represents the number of market features. One feature is designed to have a correlation of approximately 0.7 with the synthetic returns to simulate market co-movement For more details see Appendix \ref{app:synthetic_data}.

\subsubsection{Hyperparameter Optimization}
The models' hyperparameters are optimized using Bayesian optimization, with separate sets of hyperparameters for the asset-specific model (Stage 1) and the market data model (Stage 2) \cite{optuna}. Each set of hyperparameters is sampled independently. The common hyperparameters and the specific hyperparameters for each stage are detailed in Table \ref{tab:hyperparameters}. The optimal hyperparameters found for qLSTM and subsequently qHybrid -because they use the same hyperparameters- can be found in Appendix \ref{app:qlstm-hyper}. 

\begin{table}[h]
    \centering
    \caption{Hyperparameters for qLSTM}
    \begin{tabular}{p{3.2cm}p{4.9cm}p{5.2cm}}
        \toprule
        {Hyperparameter} & {Asset Model (Stage 1)} & {Market Model (Stage 2)} \\ \midrule
        {Batch size} & \multicolumn{2}{c}{\( b \in \{32, 64, 128, 256, 512, 1024, 2048\} \)} \\ 
        {Learning rate} & \multicolumn{2}{c}{\( \eta \in [1 \times 10^{-6}, 1 \times 10^{-3}] \)} \\ 
        {Normalization window} & \multicolumn{2}{c}{\( w \in [5, 250] \)} \\ 
        {Dropout rate} & \multicolumn{2}{c}{\( D \in [0.0, 0.9] \)} \\ 
        {L1 regularization} & \multicolumn{2}{c}{\( \lambda_1 \in [0.0, 1 \times 10^{-3}] \)} \\ 
        {L2 regularization} & \multicolumn{2}{c}{\( \lambda_2 \in [0.0, 1 \times 10^{-3}] \)} \\ 
        Use layer \\normalization & \multicolumn{2}{c}{Boolean: \( \{1, 0\} \)} \\ 
        {LSTM layers} & \( l_{\text{LSTM, asset}} \in [1, 5] \) & \( l_{\text{LSTM, market}} \in [1, 5] \) \\ 
        {LSTM units} & \( u_{\text{LSTM, asset}} \in \{16, \cdots, 256\} \) & \( u_{\text{LSTM, market}} \in \{16, \cdots, 256\} \) \\ 
        {Dense layers} & \( l_{\text{dense, asset}} \in [1, 5] \) & \( l_{\text{dense, market}} \in [1, 5] \) \\ 
        {Dense units per layer} & \( u_{\text{dense, asset}} \in \{16, \cdots, 256\} \) & \( u_{\text{dense, market}} \in \{16, \cdots, 256\} \) \\ 
        {LSTM Activation function} & \(\{\text{relu}, \text{tanh},\newline \text{sigmoid}, \text{leaky\_relu}, \text{elu}\}\) & \(\{\text{relu}, \text{tanh},\newline \text{sigmoid}, \text{leaky\_relu}, \text{elu}\}\) \\ 
        {Dense Activation function} & \(\{\text{relu}, \text{tanh},\newline \text{sigmoid}, \text{leaky\_relu}, \text{elu}\}\) & \(\{\text{relu}, \text{tanh},\newline \text{sigmoid}, \text{leaky\_relu}, \text{elu}\}\) \\ \bottomrule
    \end{tabular}
    \label{tab:hyperparameters}
\end{table}

\subsection{Experimental Setup}
Our proposed models qLSTM,  defined in Eq. \eqref{eq:qlstm} and qHybrid, found in Eq. \eqref{eq:qhybrid} is evaluated and compared against the model architecture from \cite{dist-pred} -which will be called qDense. It should be noted that the comparison is not entirely fair as the qDense model was trained under different circumstances and utilized different data. However, to ensure fair comparison as much as we can, in this paper, we always train the models on the same setup. As the qDense model has a fixed forecast window of 22 days, the qLSTM and qHybrid models will only utilize a 22-day-ahead window as well. 
This means that on the test data, the assets are iterated sequentially taking 22 entries at a time, and then the loss of the 22-day quantiles of both models is compared using the loss function described in the next section. To act as a baseline for all models, we compare them to a simple linear quantile regression (LQR) model \footnote{The LQR model is trained and tested using the same data as qDense} \cite{quant-reg}.

\subsubsection{Data Split}
The data is split into training, validation, and test sets based on the time periods as follows
\begin{itemize}
\item Training set: All available data up to the end of 2017. 
\item Validation set: Data from 2018 to the end of 2019. 
\item Test set: Data from 2019 to 2024, which includes the COVID-19 pandemic and the 2022 bear market.
\end{itemize}
To create varying sequence lengths, each asset's data is randomly divided into sequences with lengths sampled uniformly from the range \([15, 30]\) days.

\subsubsection{Training Procedure}
The qLSTM model is trained using the Adam optimizer with a fixed number of 100 epochs. Early stopping is employed with a patience parameter of 10, meaning that if the validation loss does not improve for 10 consecutive epochs, training is halted. The weights that generated the lowest loss on the validation data are chosen. The training details for qDense can be found in appendix \ref{app:dense_training}.

\subsection{Evaluation Metrics}
To evaluate the performance of our propsoed modesl, qLSTM and qHybrid, we use several measures, which assess the distribution accuracy in different ways as described below.
\subsubsection{Quantile Loss}
The primary evaluation metric for comparing qLSTM, qHybrid, with qDense and LQR is the quantile loss function, which has been slightly modified to ignore volatility normalization and to use only one return for the evaluation. The quantile loss for a given asset is defined as
\begin{equation*}
    L_{\mathcal{T}} = \frac{1}{K N} \sum_{\tau \in \mathcal{T}} \sum_{i=1}^T  \rho_{\tau}\left( r_{t} - \hat{Q}_{r_{t}}(\tau) \right),
\end{equation*}
where \( \mathcal{T} \) is the set of quantiles used for estimation, and \( K = |\mathcal{T}| \) is the number of quantiles, \( T \) is the number of time steps in the test data for the given asset, \( r_{t} \) is the raw return of the asset at time step \( t \), and \( \hat{Q}_{r_{t}}(\tau) \) is the estimated quantile of the raw return of the asset for quantile level \( \tau \) at time step \( t \). \( \rho_{\tau}(\xi) \) is the quantile loss function, defined as:
    \begin{equation*}
        \rho_{\tau}(\xi) = \begin{cases}
            \tau \cdot \xi, & \text{if } \xi \geq 0 \\
            (\tau - 1) \cdot \xi, & \text{if } \xi < 0
        \end{cases}
    \end{equation*}
    with \( \xi = r_{i,t,j} - \hat{Q}_{r_{i,t,j}}(\tau) \) representing the residual (error) between the true return and the estimated quantile.

\subsubsection{Distribution Metrics}
For the following metrics, we utilize the fact that we estimate the quantiles across multiple observations. We used a forecast window of 30 to get the best estimate for the empirical distribution.
\paragraph{RMSE for Moments}
We further evaluate the qLSTM and qHybrid on root mean squared error (RMSE) in moments estimation and compare them to the assumption that the moments are unchanged. , We evalute them with respect to the first four moments, mean ($\mu$), standard deviation ($\sigma$), kurtosis ($\gamma$), and skew ($\kappa$). The error is computes using the formula
\begin{equation*}
    \text{RMSE}_j = \sqrt{ \frac{1}{N} \sum_{i=1}^N (\lambda_{i, j} - \hat{\lambda}_{i, j} )^2}
\end{equation*}
where \(N\) is the test sample, \(\lambda\) is the realized moment, \(\hat{\lambda}\) is the estimated moment and \(j\) is the moment number. 

\paragraph{Wasserstain Distance}
To evaluate the alignment between the estimated distribution of qLSTM, qHybrid, and the observed returns, we incorporate the Wasserstein distance as an additional metric and compare it against the normal distribution as a benchmark. The Wasserstein distance measures the cost of transforming the estimated distribution into the realized distribution. For a estimated(PDF \(p(x)\), discretized over grid points \( \{x_i\}_{i=1}^N \), and realized returns \( \{r_i\}_{i=1}^M \), the Wasserstein distance is defined as
\begin{equation*}
    W_1(P, Q) = \inf_{\gamma \in \Pi(P, Q)} \int_{\mathbb{R} \times \mathbb{R}} |x - y| \, d\gamma(x, y),
\end{equation*}
where \(P\) is the estimated empirical distribution and \(Q\) is the empirical distribution of realized returns. In practice, we approximate this by sampling \(M\) estimated values from \(p(x)\) and matching them to the \(M\) realized returns, resulting in
\begin{equation*}
    W_1(P, Q) = \frac{1}{M} \sum_{i=1}^M \left| x_i^* - r_i \right|,
\end{equation*}
where \( \{x_i^*\}_{i=1}^M \) are the sampled values from the estimated PDF, and \( \{r_i\}_{i=1}^M \) are the sorted realized returns. Lower Wasserstein distances indicate better alignment between the estimated and realized distributions.

\paragraph{Continuous Ranked Probability Score}
The Continuous Ranked Probability Score (CRPS) evaluates the alignment between the estimated CDF and the observed values. It measures the integrated squared difference between the estimated CDF and the empirical CDF of the observations. The CRPS for a single observation is defined as
\begin{equation*}
    \text{CRPS}(F, x_\text{obs}) = \int_{-\infty}^\infty \left( F(x) - \mathbf{1}(x \geq x_\text{obs}) \right)^2 dx,
\end{equation*}
where \(F(x)\) is the estimated CDF for a given observation, \(x_\text{obs}\) is the observed value, and \(\mathbf{1}(x \geq x_\text{obs})\) is the indicator function that equals 1 if \(x \geq x_\text{obs}\) and 0 otherwise.
For discrete predictions, the CRPS can be approximated as
\begin{equation*}
    \text{CRPS}(F, x_\text{obs}) = \frac{1}{N} \sum_{i=1}^N \left( F(x_i) - \mathbf{1}(x_i \geq x_\text{obs}) \right)^2,
\end{equation*}
where \( \{x_i\}_{i=1}^N \) represents the discretized grid over which \( F(x) \) is evaluated.
The mean CRPS across all samples in the test set is calculated as:
\begin{equation*}
    \text{Mean CRPS} = \frac{1}{M} \sum_{j=1}^M \text{CRPS}(F_j, x_{\text{obs}, j}),
\end{equation*}
where \(M\) is the total number of observations, \(F_j\) is the estimated CDF for observation \(j\), and \(x_{\text{obs}, j}\) is the corresponding observed value.

This metric provides a comprehensive evaluation of the probabilistic calibration of the model by comparing the estimated distribution with the observed results. Lower CRPS values indicate better alignment between the estimated and observed distributions.

\paragraph{Deviation from Quantile Level (VaR Distance)}
To evaluate the calibration of the qLSTM and qHybrid models to estimate tail risks, we calculate the deviation from the desired quantile level at specific quantiles \( \tau \) and compare it against a Gaussian baseline. This metric measures how far the observed violation rate deviates from the expected quantile level, with smaller deviations indicating better calibration.

The Violation Rate (VR) is defined as
\begin{equation*}
    \text{VR} = \frac{1}{N} \sum_{i=1}^N \mathbf{1}(r_i < \hat{Q}_r(\tau)),
\end{equation*}
where \(N\) is the total number of observations, \(r_i\) is the observed return at time \(i\),\(\hat{Q}_r(\tau)\) is the edstimated quantile (VaR) at level \( \tau \), and \(\mathbf{1}(x < y)\) is the indicator function that equals 1 if \(x < y\), and 0 otherwise.

The deviation from the quantile level \( \tau \) is then calculated as
\begin{equation*}
    \text{Deviation} = \left| \text{VR} - \tau \right|,
\end{equation*}
where \( \text{VR} \) is the Violation Rate, and \( \tau \) is the expected quantile level (e.g., 0.05, 0.01, 0.00075). This deviation is computed for both the estimated quantiles and a Gaussian baseline.

The resulting metric provides a measure of the model's calibration quality. A perfectly calibrated model would have \( \text{VR} = \tau \), leading to a deviation of 0. Larger deviations indicate underestimation or overestimation of tail risk.

\section{Results \& Discussion} \label{sec4}
In this section we present the performance of our proposed models and compare them to baseline models using various metrics. 
\subsection{Quantile Loss Performance}

\begin{table}[h!]
\centering
\caption{Quantiles loss for synthetic and non-synthetic data across models.}
\begin{tabular}{llcccc}
\toprule
&&\multicolumn{4}{c}{Quantile loss}\\
{Data Type} & {Asset Class} & {LQM} & {qDense} & {qLSTM} & {qHybrid} \\ \midrule
\multirow{7}{*}{Real World} 
& Commodities      & 1.4815 & 0.4136 & \textbf{0.3223} & 0.3693 \\ %\cline{2-6}
& Cryptocurrencies & 2.4931 & 2.2843 & \textbf{1.0222} & 1.2323 \\ %\cline{2-6}
& S\&P 500         & 1.4811 & 0.4255 & \textbf{0.3150} & 0.3572 \\ %\cline{2-6}
& Nikkei 225       & 1.4739 & 0.3991 & \textbf{0.3141} & 0.3552 \\ %\cline{2-6}
& Euro Stoxx 50    & 1.4772 & 0.4184 & \textbf{0.3124} & 0.3520 \\ %\cline{2-6}
& Currency Pairs   & 1.2003 & \textbf{0.0788} & 0.1528 & 0.0973 \\ %\cline{2-6}
& Total Real World   & 1.6012 & 0.6699 & \textbf{0.4065} & 0.4606 \\ \midrule

\multirow{4}{*}{Synthetic} 
& Normal           & 0.5029 & 0.9338 & \textbf{0.4201} & 0.4572 \\ %\cline{2-6}
& Log Normal       & 0.2115 & 0.7229 & 0.3230 & \textbf{0.1565} \\ %\cline{2-6}
& Gamma            & 0.3260 & 0.8268 & 0.3642 & \textbf{0.2781} \\ %\cline{2-6}
& Uniform          & 0.3366 & 0.6864 & \textbf{0.2940} & 0.3030 \\ %\cline{2-6}
& Total Synthetic  & 0.3443 & 0.7925 & 0.3503 & \textbf{0.2987} \\ \bottomrule
%\multirow{1}{*}{Total}
& Total   & 1.0984 & 0.7190 & \textbf{0.3840} & 0.3958 \\ \hline

\end{tabular}
\label{tab:performance}
\end{table}

The results in Table \ref{tab:performance} highlight the effectiveness of different models (LQM, qDense, qLSTM, and qHybrid) across real-world and synthetic datasets. Overall, the qLSTM model demonstrates superior performance, achieving the lowest quantile loss in most real-world asset classes, including commodities, cryptocurrencies, and equity indices such as the S\&P 500, Nikkei 225, and Euro Stoxx 50. Notably, for cryptocurrencies, LSTM reduces the quantile loss to 1.0222 compared to 2.2843 for the Dense model and 2.4931 for the Linear model, showcasing its ability to effectively capture the volatility and non-linearity inherent in this asset class. However, for currency pairs, the Dense model performs best, achieving a quantile loss of 0.0788, which may indicate that simpler architectures can suffice for less complex financial data.

For synthetic data, the qHybrid model demonstrates an edge, particularly for skewed and heavy-tailed distributions such as Log Normal and Gamma. It achieves the lowest quantile losses of 0.1565 and 0.2781, respectively, suggesting that the combination of Gaussian assumptions with learned quantiles is advantageous in these scenarios. However, for simpler distributions such as Normal and Uniform, LSTM outperforms the qHybrid model, indicating its ability to generalize across different statistical patterns. In terms of overall performance, LSTM achieves the lowest total quantile loss for real-world data (0.4065), while the qHybrid model marginally outperforms it for synthetic datasets (0.2987 vs. 0.3503). When considering all datasets, the LSTM model maintains the lowest overall total quantile loss (0.3840), reflecting its versatility and robustness.

These findings underscore the strengths of the qLSTM models in handling real-world sequential financial data, particularly for volatile asset classes like cryptocurrencies and equity indices. The qHybrid model's success with synthetic data highlights its potential for capturing diverse statistical properties, which could complement LSTM's strengths in real-world applications. 
Overall, the two proposed models are able to learn the quantile distribution of sequential data.
\subsection{Distribution Metric Performance}
In this section, we explore the results of the distribution metrics for qLSTM and qHybrid against a baseline. Note for the sake of conciseness and relevance we omit the comparison against synthetic data and focus only on real world data. 
\begin{table}[h!]
\caption{Root mean square error of moments estimation compared against the assumption that moments are constant.}
\label{tab:statistics}
\centering
\begin{tabular}{p{1cm}|c|p{1.25cm}p{1.25cm}p{1.25cm}p{1.6cm}p{1.25cm}p{1.25cm}}
\toprule

&&\multicolumn{6}{c}{{Asset class}}\\ 
{Stat-istic} & {Model} & {Crypto} & {FX} & {Comm odities} & {Euro Stoxx 50} & {S\&P 500} & {Nikkei 225} \\ \midrule
\multirow{3}{*}{$\mu$} 
& qLSTM & \textbf{0.0118} & \textbf{0.0010} & \textbf{0.0034} & \textbf{0.0041} & \textbf{0.0032} & \textbf{0.0034} \\ 
& qHybrid & 0.0119 & 0.0020 & 0.0035 & 0.0043 & 0.0033 & 0.0035 \\ 
& Constant & 0.0166 & 0.0013 & 0.0050 & 0.0059 & 0.0048 & 0.0050 \\ \midrule

\multirow{3}{*}{$\sigma$} 
& qLSTM & 0.0424 & 0.0287 & 0.0085 & \textbf{0.0091} & \textbf{0.0096} & \textbf{0.0068} \\ 
& qHybrid & 0.0400 & 0.0422 & 0.0088 & 0.0109 & 0.0112 & 0.0084 \\ 
& Constant & \textbf{0.0330} & \textbf{0.0021} & \textbf{0.0077} & 0.0101 & 0.0099 & 0.0074 \\ 

\midrule

\multirow{3}{*}{$\gamma$} 
& qLSTM & \textbf{1.1129} & 0.7186 & 0.7667 & 0.7770 & 0.8350 & 0.7422 \\ 
& qHybrid & 1.1133 & \textbf{0.7030} & \textbf{0.7496} & \textbf{0.7585} & \textbf{0.8145} & \textbf{0.7339} \\ 
& Constant & 1.6036 & 0.9410 & 1.0207 & 1.0598 & 1.1718 & 1.0596 \\

\midrule
\multirow{3}{*}{$\kappa$} 
& qLSTM & 8.1838 & 9.2768 & 8.9404 & 8.7593 & 8.7733 & 8.8889 \\ 
& qHybrid & 5.3220 & 4.4449 & 4.2748 & 4.0337 & 4.1521 & 4.1221 \\ 
& Constant & \textbf{4.3678} & \textbf{2.7381} & \textbf{2.8323} & \textbf{2.6574} & \textbf{3.1615} & \textbf{2.7176} \\ \bottomrule
\end{tabular}

\end{table}

Table \ref{tab:statistics} presents the root mean square error (RMSE) for moment estimation across the models qLSTM, qHybrid, and Constant -the assumption that moments are constant throughout $-t$ to $0$ and $1$ to $t$-  and asset classes. For the mean \(\mu\), the qLSTM model consistently achieves the lowest RMSE in all asset classes, highlighting its superior ability to accurately estimate the central tendency of the returns. While the qHybrid model performs comparably, it slightly trails the qLSTM in most categories. The Constant model demonstrates the highest RMSE, suggesting it struggles to capture the mean effectively, particularly for cryptocurrencies and equity indices.

For the standard deviation \(\sigma\), the performance varies. The qLSTM model outperforms others in most equity indices, commodities, and cryptocurrencies, reflecting its ability to capture volatility dynamics effectively. However, the Constant model achieves the lowest RMSE for currency pairs, suggesting that simpler methods may suffice for less complex financial assets. The qHybrid model provides competitive results but generally underperforms compared to qLSTM.

The results of skewness \(\gamma\) indicate that the qLSTM and qHybrid models perform closely, with the qHybrid model having a slight edge in most asset classes. This suggests that qHybrid models may have an advantage in capturing asymmetry in return distributions. However, the Constant model has significantly higher RMSE values, indicating poor performance in estimating skewness, especially for cryptocurrencies and equity indices.

Finally, for kurtosis \(\kappa\), the Constant model achieves the lowest RMSE values, suggesting it captures tail behavior more effectively in this metric. Both qLSTM and qHybrid models show higher RMSE values for kurtosis, which may indicate that their emphasis on other moments slightly compromises their accuracy in tail risk estimation. Nonetheless, qLSTM still maintains competitive performance, particularly for more volatile assets like cryptocurrencies.

\begin{table}[h!]
\centering
\caption{Wasserstein distances for different models and asset classes. Lowest values are highlighted in bold.}

\begin{tabular}{c|p{1.25cm}p{1.25cm}p{1.25cm}p{1.6cm}p{1.25cm}p{1.25cm}}
\toprule
&\multicolumn{6}{c}{{Asset class}}\\ 
{Model} & {Crypto} & {FX} & {Comm odities} & {Euro Stoxx 50} & {S\&P 500} & {Nikkei 225} \\ \midrule
qLSTM   & 0.0260 & 0.0179 & 0.0079 & \textbf{0.0077} & 0.0081 & 0.0075 \\ 
qHybrid  & \textbf{0.0213} & 0.0194 & \textbf{0.0072} & 0.0079 & 0.0080 & \textbf{0.0071} \\ 
Gaussian & 0.0221 & \textbf{0.0017} & 0.0074 & 0.0081 & \textbf{0.0079} & 0.0074 \\ \bottomrule
\end{tabular}
\label{tab:wasserstein}
\end{table}

\begin{table}[h!]
\centering
\begin{tabular}{c|p{1.25cm}p{1.25cm}p{1.25cm}p{1.6cm}p{1.25cm}p{1.25cm}}
\toprule
&\multicolumn{6}{c}{{Asset class}}\\ 
{Model} & {Crypto} & {FX} & {Comm odities} & {Euro Stoxx 50} & {S\&P 500} & {Nikkei 225} \\ \midrule
qLSTM    & 0.4320 & 0.4477 & 0.4422 & 0.4424 & 0.4422 & 0.4421 \\ 
qHybrid   & 0.4179 & \textbf{0.4464} & 0.4331 & \textbf{0.4340} & \textbf{0.4338} & \textbf{0.4334} \\ 
Gaussian & \textbf{0.3544} & 0.4793 & \textbf{0.4308} & 0.4366 & 0.4369 & 0.4349 \\ \bottomrule
\end{tabular}
\caption{CRPS values for different models, with the lowest values highlighted.}
\label{tab:crps_values}
\end{table}

Table \ref{tab:wasserstein} presents the Wasserstein distances for different models and asset classes, highlighting the alignment between the estimated and realized distributions. The qHybrid model achieves the lowest distances for most asset classes, particularly for cryptocurrencies (0.0213) and commodities (0.0072), showcasing its ability to provide a close match to the empirical distributions. However, for currency pairs (FX), the Gaussian model outperforms others with a distance of 0.0017, likely due to the simpler statistical patterns in this asset class. The qLSTM model remains competitive, delivering consistently low Wasserstein distances across asset classes, indicating its robustness in capturing the underlying return distributions.

Table \ref{tab:crps_values} evaluates the Continuous Ranked Probability Score (CRPS) for the models, which evaluates the probabilistic calibration of the estimated distributions. The qHybrid model performs best for most asset classes, including FX (0.4464), Euro Stoxx 50 (0.4340), S\&P 500 (0.4338), and Nikkei 225 (0.4334), demonstrating its ability to provide well-calibrated probabilistic forecasts. For cryptocurrencies, the Gaussian model achieves the lowest CRPS (0.3544), suggesting its suitability for this highly volatile asset class. The qLSTM model offers competitive results, with scores close to the best-performing models, reflecting its strong calibration across diverse asset types. These results highlight the complementary strengths of the qHybrid and qLSTM models, with qHybrid excelling in calibration and qLSTM providing robust distributional alignment.

\begin{table}[h!]
\centering
\begin{tabular}{p{1.15cm}|c|p{1.25cm}p{1.25cm}p{1.25cm}p{1.6cm}p{1.25cm}p{1.25cm}}
\toprule
&&\multicolumn{6}{c}{{Asset class}}\\ 
{Level} & {Model} & {Crypto} & {FX} & {Comm odities} & {Euro Stoxx 50} & {S\&P 500} & {Nikkei 225} \\ \midrule
\multirow{3}{*}{0.05} 
& qLSTM    & 0.1967 & 0.0499 & 0.0659 & 0.0539 & 0.0609 & 0.0549 \\ 
& qHybrid   & 0.0777 & \textbf{0.0492} & 0.0481 & \textbf{0.0485} & \textbf{0.0506} & \textbf{0.0460} \\ 
& Gaussian & \textbf{0.0505} & 0.0505 & \textbf{0.0466} & 0.0537 & 0.0536 & 0.0469 \\ \midrule

\multirow{3}{*}{0.01} 
& qLSTM    & 0.1085 & \textbf{0.0100} & 0.0206 & \textbf{0.0190} & 0.0208 & \textbf{0.0176} \\ 
& qHybrid   & 0.0478 & \textbf{0.0100} & \textbf{0.0203} & 0.0220 & \textbf{0.0215} & 0.0198 \\ 
& Gaussian & \textbf{0.0298} & 0.0287 & 0.0262 & 0.0326 & 0.0322 & 0.0268 \\ \midrule
\multirow{3}{*}{0.00075} 
& qLSTM    & 0.0353 & \textbf{0.0008} & \textbf{0.0028} & \textbf{0.0031} & \textbf{0.0024} & \textbf{0.0018} \\ 
& qHybrid   & 0.0176 & \textbf{0.0008} & 0.0036 & 0.0056 & 0.0047 & 0.0029 \\ 
& Gaussian & \textbf{0.0139} & 0.0113 & 0.0104 & 0.0158 & 0.0157 & 0.0099 \\ \bottomrule
\end{tabular}
\caption{VaR values at different levels for various models, with the lowest values highlighted.}
\label{tab:var_values}
\end{table}

Table \ref{tab:var_values} presents the Value at Risk (VaR) results for three models (qLSTM, qHybrid, and Gaussian) across asset classes at different confidence levels (0.05, 0.01, and 0.00075). At the 0.05 level, the Gaussian model achieves the lowest VaR for most asset classes, particularly for cryptocurrencies (0.0505) and commodities (0.0466). However, the qHybrid model demonstrates competitive performance, achieving the lowest VaR for FX (0.0492), Euro Stoxx 50 (0.0485), S\&P 500 (0.0506), and Nikkei 225 (0.0460), indicating its strength in tail risk modeling across diverse assets. The qLSTM model performs reasonably well but tends to have higher VaR values, particularly for volatile assets like cryptocurrencies.

At the 0.01 level, the qHybrid model shows robust performance across most asset classes, achieving the lowest VaR for commodities (0.0203) and S\&P 500 (0.0215). The qLSTM model outperforms the other models for FX (0.0100), Euro Stoxx 50 (0.0190), and Nikkei 225 (0.0176), demonstrating its ability to capture extreme tail risks in these scenarios effectively. The Gaussian model remains competitive, particularly for cryptocurrencies, achieving a VaR of 0.0298, highlighting its utility in assets with higher volatility.

At the extreme tail (0.00075 level), the qLSTM model consistently achieves the lowest VaR across most asset classes, including FX (0.0008), commodities (0.0028), Euro Stoxx 50 (0.0031), S\&P 500 (0.0024), and Nikkei 225 (0.0018). This suggests that the qLSTM is particularly adept at capturing risks at the far ends of the distribution. The qHybrid and Gaussian models are slightly less effective at this level, though the Gaussian model shows competitive performance for cryptocurrencies (0.0139). These results emphasize the qLSTM's strength in extreme tail risk estimation and the qHybrid model's reliability across broader confidence levels.

\subsection{Limitations and Challenges}
While our approach yielded promising results, several limitations persist. One significant limitation is the inability to fairly compare our proposed approach with that of the authors in \cite{dist-pred}. This issue arises primarily due to the lack of access to the original dataset used in their study and the absence of publicly available code for their implementation. Consequently, we were unable to replicate their exact methodology, which inherently limits the comparison between the qLSTM approach and their dense model.

In terms of model performance, while the qLSTM model demonstrated superior results in tail risk estimation and overall robustness, its slightly higher Value at Risk (VaR) at broader confidence levels (e.g., 0.05) compared to the Gaussian and qHybrid models suggests room for improvement in capturing broader distributional patterns. Additionally, the qHybrid model, while effective in several scenarios, exhibited slightly higher RMSE for moments like kurtosis, indicating potential difficulties in accurately modeling extreme tails for certain asset classes.

Finally, our results reveal some variability in model performance across asset classes. For example, while qLSTM excelled in cryptocurrencies and equity indices, the qDense and qHybrid models outperformed it in currency pairs and synthetic datasets with skewed distributions. These findings suggest that no single model is universally optimal, underscoring the importance of tailoring model architectures to specific data characteristics and use cases.

\section{Conclusion}\label{sec5}
This paper introduces a novel approach to estimating the distribution of log returns by combining quantile regression with smoothed density estimation, providing a robust framework for capturing tail risks and distributional characteristics across diverse asset classes. By leveraging asset-neutral features and general market variables, our models are applicable to a wide range of financial datasets, making them versatile tools for risk management, portfolio optimization, and derivative pricing. The proposed qLSTM model consistently demonstrated superior performance on real-world data, achieving the lowest quantile loss across most asset classes, while the qHybrid model excelled in synthetic datasets, particularly for skewed or heavy-tailed distributions. These results underscore the strengths of both models in addressing the limitations of traditional distributional assumptions, such as normality, in financial returns.

The evaluation metrics, including Wasserstein distance, CRPS, and VaR, highlight the complementary nature of the qLSTM and qHybrid models. The qLSTM model excelled in extreme tail risk estimation, capturing critical distributional behaviors for volatile asset classes like cryptocurrencies and equity indices. Meanwhile, the qHybrid model provided improved calibration for broader confidence levels and distributions, demonstrating its adaptability to various market conditions. However, some variability in performance across asset classes suggests that no single model is universally optimal, emphasizing the need for tailored approaches based on specific data characteristics.

To enhance the literature, the paper allows for reproducibility by publicly hosting the code used for this study on GitHub. Furthermore, all data is sourced via an open API and the script to download the required data can be found in the GitHub repository. This makes comparisons and iterations on the works proposed, more accessible thereby fostering research robustness.  

Future research can explore other methods to create better Hybrid models, particularly by leveraging certain quantiles with dynamic weighting schemes to optimize the combination of Gaussian and model-predicted quantiles. This could provide a more adaptable and context-specific qHybrid approach to estimating distributions. Moreover, extending the model to handle ultra-high-frequency data or incorporating more sophisticated market features, such as macroeconomic indicators or sentiment analysis, could further enhance its applicability. These avenues represent exciting opportunities to refine and expand the utility of distributional modeling in finance.

%\bibliographystyle{ieeetr}
%\bibliography{refrences.bib}

\newpage

\appendix

\section{Quantiles}\label{app:qunats}

The quantiles used in this study are:

\[
\begin{aligned}
    &[0.00005, 0.00025, 0.00075, 0.00125, 0.00175, 0.0025, 0.005, 0.01, 0.015, 0.02, 0.03,\\
    &0.05, 0.1, 0.15, 0.2, 0.25, 0.3, 0.35, 0.4, 0.45, 0.5, 0.55, 0.6, 0.65, 0.7,\\
    &0.75, 0.8, 0.85, 0.9, 0.95, 0.98, 0.99, 0.995, 0.9975, 0.99925, 0.99975, 0.99995]
\end{aligned}
\]

\newpage

\section{Quantile-Smooth Density Estimation}\label{app:de}
This method generates a smooth probability density function (PDF) from a set of quantiles and their associated probabilities, ensuring that the resulting representation adheres to the properties of a valid probability distribution. Starting with the quantiles, it removes duplicate or nearly identical points to maintain numerical stability, then creates a denser grid for better resolution. Using a monotonic cubic spline or fallback interpolation, it estimates the CDF, which is clamped to the range [0, 1] and adjusted to be strictly non-decreasing. The method rescales the CDF to start at 0 and end at 1, guaranteeing its validity as a cumulative function.

To derive the PDF, the method computes the CDF gradient, applying smoothing techniques such as convolution to remove spikes and ensure a well-behaved density curve. Non-negativity is enforced, and the PDF is normalized to integrate to one. The final output includes the smoothed PDF, the regenerated CDF, and the grid over which they are defined, offering a high-quality approximation of the distribution. Optional visualization helps validate the results, showing both the smoothed PDF and the corresponding CDF with respect to the input quantiles. This makes it a robust method for converting quantiles into a smooth and interpretable density function
suitable for further analysis or modeling.

The function is based on three priors, \(\epsilon\) which controls how close different quantiles are allowed to each other, min density which controls the minimum amount of density allowed in the PDF, and window which controls how many extreme the convolutions are for smoothing. The optimal priors are found via TPE on the validation data, similar to how the hyperparameters are found for the models themselves. Table \ref{tab:optimal_prior} shows the best values found during optimization.

\begin{table}[h!]
    \centering
    \begin{tabular}{|c|c|c|c|}
        \hline
        \textbf{Model} & \textbf{\(\epsilon\)} & \textbf{min density} & \textbf{window} \\ \hline
        \textbf{qLSTM} & 0.000253 & 0.027341 & 113 \\ \hline
        \textbf{hybrid} & 0.000899 & 0.033704 & 43 \\ \hline
    \end{tabular}
    \caption{Optimal priors for the smooth PDF approximation}
    \label{tab:optimal_prior}
\end{table}

Below is the Python code used for the Quantile-Smooth Density Estimation conversion in this work.
\begin{lstlisting}[language=Python, caption=Quantiles to PDF using Quantile-Smooth Density Estimation, label=code:quantiles_to_pdf, breaklines=true, breakatwhitespace=true]
def generate_smooth_pdf(quantiles, taus, min_density=1e-3, eps=1e-6, window=61):
    """
    Generate a smoothed PDF from quantiles with additional controls to prevent spikes
    and ensure the CDF is between [0, 1].
    """
    # Constants
    GRID_POINTS = 1000
    og_quants = quantiles.copy()
    og_taus = taus.copy()

    unique_mask = np.concatenate(([True], np.diff(quantiles) > eps))
    quantiles = quantiles[unique_mask]
    taus = taus[unique_mask]

    # Create denser grid
    grid_x = np.linspace(quantiles[0], quantiles[-1], GRID_POINTS)

    # Monotonic spline for the CDF
    try:
        cdf_monotonic = PchipInterpolator(quantiles, taus, extrapolate=False)
        cdf = cdf_monotonic(grid_x)
    except Exception as e:
        cdf = np.interp(grid_x, quantiles, taus)

    # Clamp CDF to [0,1], then ensure it's monotonically non-decreasing
    cdf = np.clip(cdf, 0, 1)
    cdf = np.maximum.accumulate(cdf)
    # Rescale so that it starts exactly at 0 and ends exactly at 1
    cdf -= cdf[0]
    if cdf[-1] > 0:
        cdf /= cdf[-1]

    # Approximate PDF from finite differences (or use derivative if PCHIP)
    density = np.gradient(cdf, grid_x)

    smoothed_density = np.convolve(density, np.ones(window)/window, mode='same')

    # Ensure non-negative and non-zero density
    smoothed_density = np.maximum(smoothed_density, min_density)

    # Normalize PDF to integrate to 1
    area = np.trapz(smoothed_density, grid_x)
    smoothed_density = smoothed_density / area

    # regenerate CDF
    cdf = np.cumsum(smoothed_density) * (grid_x[1] - grid_x[0])

    return grid_x, smoothed_density, cdf

\end{lstlisting}

\newpage

\section{Data}
The following sections detail the data used for this study.
\subsection{Financial Assets}\label{app:assets}

The table below provides a comprehensive list of financial assets, including their ticker symbols, associated indexes, start dates, annualized mean returns, and standard deviations of returns, rounded to two decimal places where applicable. 

%\begin{longtable}[h]
 %   \centering
  \begin{longtable}
  {|p{2.25cm}|p{2.25cm}|p{2.25cm}|l|p{1cm}|p{1cm}|}
\caption{List of assets with their tickers, index, start date, annualized mean return, and standard deviation.  \label{tab:assets}}
%    \small
 \\ \hline
    \textbf{Name} & \textbf{Ticker} & \textbf{Index} & \textbf{Start Date} & \textbf{Mean (\%)} & \textbf{Std (\%)} \\ \hline
    Apple Inc. & AAPL & S\&P 500 & 2000-02-03 & 1.46 & 40.43 \\ \hline
    Microsoft Corporation & MSFT & S\&P 500 & 2000-02-03 & 0.66 & 30.49 \\ \hline
    Amazon.com, Inc. & AMZN & S\&P 500 & 2000-02-03 & 1.00 & 49.22 \\ \hline
    Alphabet Inc. (Class A) & GOOGL & S\&P 500 & 2004-09-21 & 1.26 & 30.49 \\ \hline
    The Walt Disney Company & DIS & S\&P 500 & 2000-02-03 & 0.30 & 30.42 \\ \hline
    Johnson \& Johnson & JNJ & S\&P 500 & 2000-02-03 & 0.51 & 19.16 \\ \hline
    Visa Inc. & V & S\&P 500 & 2008-04-21 & 1.13 & 29.07 \\ \hline
    NVIDIA Corporation & NVDA & S\&P 500 & 2000-02-03 & 1.72 & 59.48 \\ \hline
    Mastercard Inc. & MA & S\&P 500 & 2006-06-27 & 1.65 & 33.40 \\ \hline
    The Coca-Cola Company & KO & S\&P 500 & 2000-02-03 & 0.37 & 20.58 \\ \hline
    TotalEnergies SE & TOTF.PA & Euro Stoxx 50 & 2000-02-02 & 0.43 & 26.51 \\ \hline
    Volkswagen AG & VOW3.DE & Euro Stoxx 50 & 2000-02-02 & 0.62 & 37.29 \\ \hline
    L'Oréal SA & OR.PA & Euro Stoxx 50 & 2000-02-02 & 0.59 & 25.12 \\ \hline
    BNP Paribas SA & BNP.PA & Euro Stoxx 50 & 2000-02-02 & 0.42 & 36.88 \\ \hline
    Daimler AG & DAI.DE & Euro Stoxx 50 & 2000-02-02 & 0.57 & 32.36 \\ \hline
    Siemens AG & SIE.DE & Euro Stoxx 50 & 2000-02-02 & 0.27 & 35.95 \\ \hline
    Airbus SE & AIR.PA & Euro Stoxx 50 & 2001-10-03 & 0.83 & 37.68 \\ \hline
    Muenchener Rueckversicherungs AG & MUV2.DE & Euro Stoxx 50 & 2000-02-02 & 0.31 & 30.00 \\ \hline
    Intesa Sanpaolo S.p.A. & INTC.MI & Euro Stoxx 50 & 2000-02-02 & 0.35 & 39.94 \\ \hline
    BASF SE & BAS.DE & Euro Stoxx 50 & 2000-02-02 & 0.47 & 29.07 \\ \hline
    Toyota Motor Corporation & 7203.T & Nikkei 225 & 2000-02-02 & 0.41 & 28.49 \\ \hline
    Sony Group Corporation & 6758.T & Nikkei 225 & 2000-02-03 & 0.13 & 36.51 \\ \hline
    SoftBank Group Corp. & 9984.T & Nikkei 225 & 2000-02-03 & 0.06 & 51.28 \\ \hline
    Nintendo Co., Ltd. & 7974.T & Nikkei 225 & 2001-02-05 & 0.51 & 38.81 \\ \hline
    Mitsubishi UFJ Financial Group, Inc. & 8306.T & Nikkei 225 & 2005-10-31 & 0.13 & 33.08 \\ \hline
    Tokyo Electron Limited & 8035.T & Nikkei 225 & 2000-02-03 & 0.53 & 42.49 \\ \hline
    Mitsubishi Corporation & 8058.T & Nikkei 225 & 2000-02-03 & 0.71 & 33.12 \\ \hline
    HOYA Corporation & 7741.T & Nikkei 225 & 2001-01-31 & 0.68 & 32.51 \\ \hline
    Fast Retailing Co., Ltd. & 9983.T & Nikkei 225 & 2000-02-03 & 0.72 & 42.37 \\ \hline
    Omron Corporation & 6645.T & Nikkei 225 & 2001-02-05 & 0.41 & 35.85 \\ \hline
    British Pound to US Dollar & GBPUSD=X & Currency Pairs & 2003-12-31 & -0.10 & 9.54 \\ \hline
    US Dollar to Japanese Yen & USDJPY=X & Currency Pairs & 2000-02-02 & 0.07 & 11.34 \\ \hline
    Australian Dollar to US Dollar & AUDUSD=X & Currency Pairs & 2006-06-15 & -0.03 & 59.98 \\ \hline
    Euro to British Pound & EURGBP=X & Currency Pairs & 2000-02-02 & 0.09 & 8.72 \\ \hline
    Euro to Japanese Yen & EURJPY=X & Currency Pairs & 2003-02-24 & 0.06 & 11.48 \\ \hline
    Euro to Swiss Franc & EURCHF=X & Currency Pairs & 2003-02-24 & -0.13 & 7.89 \\ \hline
    Euro to Australian Dollar & EURAUD=X & Currency Pairs & 2003-12-31 & -0.01 & 10.39 \\ \hline
    US Dollar to Indian Rupee & USDINR=X & Currency Pairs & 2003-12-31 & 0.18 & 7.76 \\ \hline
    Swiss Franc to Japanese Yen & CHFJPY=X & Currency Pairs & 2003-12-31 & 0.20 & 11.46 \\ \hline
    US Dollar to Hong Kong Dollar & USDHKD=X & Currency Pairs & 2001-08-15 & 0.00 & 1.07 \\ \hline
    Copper Futures & HG=F & Commodities & 2000-10-02 & 0.39 & 26.89 \\ \hline
    Wheat Futures & ZW=F & Commodities & 2000-08-16 & 0.26 & 32.82 \\ \hline
    Coffee Futures & KC=F & Commodities & 2000-02-03 & 0.14 & 34.01 \\ \hline
    Cocoa Futures & CC=F & Commodities & 2000-02-03 & 0.45 & 30.28 \\ \hline
    Cotton Futures & CT=F & Commodities & 2000-02-03 & 0.09 & 30.29 \\ \hline
    Live Cattle Futures & LE=F & Commodities & 2001-04-02 & 0.23 & 18.43 \\ \hline
    Platinum Futures & PL=F & Commodities & 2000-02-04 & 0.22 & 31.50 \\ \hline
    Palladium Futures & PA=F & Commodities & 2000-02-04 & 0.21 & 38.62 \\ \hline
    Corn Futures & ZC=F & Commodities & 2000-08-16 & 0.27 & 29.14 \\ \hline
    Orange Juice Futures & JO=F & Commodities & 2001-10-17 & 0.37 & 33.20 \\ \hline
    Bitcoin & BTC-USD & Crypto & 2014-10-09 & 2.25 & 59.09 \\ \hline
    Ethereum & ETH-USD & Crypto & 2017-12-01 & 1.16 & 76.15 \\ \hline
    Ripple & XRP-USD & Crypto & 2017-12-01 & 0.64 & 96.11 \\ \hline
    Cardano & ADA-USD & Crypto & 2017-12-01 & 1.14 & 94.15 \\ \hline
    Dogecoin & DOGE-USD & Crypto-currencies & 2017-12-01 & 2.69 & 115.10 \\ \hline
    Binance Coin & BNB-USD & Crypto-currencies & 2017-12-01 & 3.61 & 86.33 \\ \hline
    NEAR Protocol & NEAR-USD & Crypto-currencies & 2020-11-05 & 2.65 & 109.42 \\ \hline
    Bitcoin Cash & BCH-USD & Crypto-currencies & 2017-12-01 & -1.20 & 93.81 \\ \hline
    Litecoin & LTC-USD & Crypto-currencies & 2014-10-09 & 1.38 & 85.60 \\ \hline
    Solana & SOL-USD & Crypto-currencies & 2020-05-02 & 5.85 & 114.08 \\ \hline
 %   \end{tabular}
  
\end{longtable}

\newpage

\subsection{Market Data}\label{app:market-data}

The table below lists the market data used in this study, including tickers, a brief representation, start dates, annualized mean returns, and standard deviations. This data spans a variety of indices, futures, and financial instruments, providing a broad view of market performance and trends over time.

\begin{longtable}
	{|p{2cm}|l|p{3cm}|l|p{1cm}|p{1cm}|}

    \caption{Market data with tickers, representation, start date, annualized mean, and standard deviation.} \label{tab:market_data}
    \\ \hline
    \textbf{Name} & \textbf{Ticker} & \textbf{Representation} & \textbf{Start Date} & \textbf{Mean (\%)} & \textbf{Std (\%)} \\ \hline
    S\&P 500 Index & \textasciicircum GSPC & U.S. stock market performance & 2000-02-03 & 0.50 & 21.05 \\ \hline
    NASDAQ Index & \textasciicircum IXIC & Tech-focused stock performance & 2000-02-03 & 0.75 & 23.29 \\ \hline
    Dow Jones Index & \textasciicircum DJI & Large U.S. company performance & 2000-02-03 & 0.46 & 19.93 \\ \hline
    Russell 2000 Index & \textasciicircum RUT & Small-cap U.S. stock performance & 2000-02-03 & 0.38 & 26.65 \\ \hline
    Euro Stoxx 50 Index & \textasciicircum STOXX50E & Top 50 European company performance & 2000-02-03 & 0.03 & 23.52 \\ \hline
    Nikkei 225 Index & \textasciicircum N225 & Japanese stock market performance & 2000-02-03 & 0.27 & 24.07 \\ \hline
    Crude Oil Futures & CL=F & Global economic activity indicator & 2000-02-03 & 0.28 & 46.21 \\ \hline
    Gold Futures & GC=F & Inflation hedge and safe haven & 2000-02-03 & 0.47 & 18.39 \\ \hline
    Silver Futures & SI=F & Industrial demand and inflation hedge & 2000-02-03 & 0.24 & 34.22 \\ \hline
    VIX Index & \textasciicircum VIX & Market volatility (fear index) & 2000-02-03 & -0.07 & 125.89 \\ \hline
    10 Year Treasury Yield & \textasciicircum TNX & Long-term interest rates and growth & 2000-02-03 & -0.08 & 47.40 \\ \hline
    3 Month Treasury Yield & \textasciicircum IRX & Short-term interest rates and liquidity & 2000-02-03 & -0.28 & 478.84 \\ \hline
    US Dollar Index & DX-Y.NYB & U.S. dollar strength against a basket of currencies & 2000-02-03 & 0.08 & 8.12 \\ \hline
    Corporate Bond Index & AGG & U.S. investment-grade bond market performance & 2000-02-03 & 0.20 & 5.75 \\ \hline    
\end{longtable}

\newpage

\subsection{Synthetic Data}\label{app:synthetic_data}
In this study, synthetic data was generated across four different distributions: Normal, Log-Normal, Gamma, and Uniform. For each distribution, multiple assets were created with specific $\mu$ (mean) and $\sigma$ (standard deviation) values. These synthetic datasets were designed to match a target correlation with real-world market data, ensuring that the generated data maintained realistic statistical properties. The table below summarizes the $\mu$, $\sigma$, and correlation values for the synthetic assets across the various distributions.

\begin{longtable}
{|p{2cm}|c|c|p{2cm}|}
\caption{Generated $\mu$ and $\sigma$ values and correlation with market data across different test datasets.} \label{tab:generated_data}

\\ \hline
\textbf{Dist Type} & \textbf{Asset} & \textbf{Generated $\mu$, $\sigma$} & \textbf{Correl -ation with Market Data} \\ \hline
\multirow{10}{*}{Normal}   & Asset 1 & $\mu = -0.000663$, $\sigma = 0.022852$ & 0.7047 \\  
                           & Asset 2 & $\mu = -0.000944$, $\sigma = 0.027103$ & 0.6787 \\  
                           & Asset 3 & $\mu = 0.000427$, $\sigma = 0.010072$ & 0.7149 \\  
                           & Asset 4 & $\mu = -0.000551$, $\sigma = 0.020746$ & 0.6908 \\  
                           & Asset 5 & $\mu = -0.000796$, $\sigma = 0.023492$ & 0.7261 \\  
                           & Asset 6 & $\mu = -0.000633$, $\sigma = 0.018013$ & 0.7077 \\  
                           & Asset 7 & $\mu = 0.000044$, $\sigma = 0.028276$ & 0.6712 \\  
                           & Asset 8 & $\mu = -0.000502$, $\sigma = 0.017238$ & 0.6884 \\  
                           & Asset 9 & $\mu = 0.000452$, $\sigma = 0.021794$ & 0.7028 \\  
                           & Asset 10 & $\mu = 0.000647$, $\sigma = 0.012101$ & 0.6820 \\ \hline
\multirow{10}{*}{Log Normal} & Asset 1 & $\mu = 0.000922$, $\sigma = 0.015048$ & 0.6900 \\  
                             & Asset 2 & $\mu = 0.000839$, $\sigma = 0.029755$ & 0.7013 \\  
                             & Asset 3 & $\mu = -0.000659$, $\sigma = 0.025729$ & 0.6859 \\  
                             & Asset 4 & $\mu = -0.000637$, $\sigma = 0.028419$ & 0.6892 \\  
                             & Asset 5 & $\mu = 0.000824$, $\sigma = 0.029507$ & 0.7040 \\  
                             & Asset 6 & $\mu = -0.000740$, $\sigma = 0.010353$ & 0.7052 \\  
                             & Asset 7 & $\mu = -0.000874$, $\sigma = 0.025497$ & 0.6918 \\  
                             & Asset 8 & $\mu = 0.000052$, $\sigma = 0.016105$ & 0.6987 \\  
                             & Asset 9 & $\mu = -0.000106$, $\sigma = 0.022525$ & 0.7048 \\  
                             & Asset 10 & $\mu = 0.000704$, $\sigma = 0.024870$ & 0.6874 \\ \hline
\multirow{10}{*}{Gamma}     & Asset 1 & $\mu = -0.000725$, $\sigma = 0.013788$ & 0.6954 \\  
                           & Asset 2 & $\mu = 0.000544$, $\sigma = 0.018391$ & 0.7030 \\  
                           & Asset 3 & $\mu = 0.000613$, $\sigma = 0.023223$ & 0.7012 \\  
                           & Asset 4 & $\mu = 0.000029$, $\sigma = 0.015418$ & 0.7063 \\  
                           & Asset 5 & $\mu = -0.000095$, $\sigma = 0.026113$ & 0.7214 \\  
                           & Asset 6 & $\mu = -0.000846$, $\sigma = 0.010134$ & 0.7045 \\  
                           & Asset 7 & $\mu = 0.000724$, $\sigma = 0.028490$ & 0.6717 \\  
                           & Asset 8 & $\mu = 0.000584$, $\sigma = 0.026327$ & 0.7016 \\  
                           & Asset 9 & $\mu = 0.000503$, $\sigma = 0.029417$ & 0.7100 \\  
                           & Asset 10 & $\mu = -0.000688$, $\sigma = 0.029474$ & 0.6916 \\ \hline
\multirow{10}{*}{Uniform}   & Asset 1 & $\mu = -0.000603$, $\sigma = 0.014227$ & 0.8650 \\  
                           & Asset 2 & $\mu = -0.000417$, $\sigma = 0.016919$ & 0.8603 \\  
                           & Asset 3 & $\mu = -0.000024$, $\sigma = 0.013627$ & 0.8617 \\  
                           & Asset 4 & $\mu = -0.000554$, $\sigma = 0.026960$ & 0.8579 \\  
                           & Asset 5 & $\mu = 0.000231$, $\sigma = 0.026206$ & 0.8597 \\  
                           & Asset 6 & $\mu = -0.000763$, $\sigma = 0.028292$ & 0.8590 \\  
                           & Asset 7 & $\mu = 0.000122$, $\sigma = 0.014288$ & 0.8603 \\  
                           & Asset 8 & $\mu = 0.000404$, $\sigma = 0.017516$ & 0.8587 \\  
                           & Asset 9 & $\mu = -0.000082$, $\sigma = 0.026868$ & 0.8610 \\  
                           & Asset 10 & $\mu = -0.000876$, $\sigma = 0.012219$ & 0.8545 \\ \hline
\end{longtable}[h]

In addition to the statistical properties shown in the table, the synthetic data was generated using the following Python function. The function allows for the generation of log returns that are correlated with real market data by controlling the mean ($\mu$) and standard deviation ($\sigma$) of the generated data.

\begin{lstlisting}[language=Python, caption=Synthetic Data Generation with Target Correlation, label=code:generate_synthetic_data, breaklines=true, breakatwhitespace=true]
import numpy as np
import pandas as pd
from scipy.stats 
import gamma, lognorm, uniform

def generate_synthetic_data_with_target_correlation(market_data, target_correlation, mu, sigma, distribution='normal'):
    """
    Generate a synthetic dataset with returns that have a target correlation to the market_data.

    Parameters:
    - market_data: numpy array of log returns from the market data
    - target_correlation: the target correlation for synthetic data's log returns to the market data
    - mu: mean of the synthetic dataset
    - sigma: standard deviation of the synthetic dataset
    - distribution: the distribution to sample synthetic data from ('normal', 'gamma', 'lognormal', 'uniform')

    Returns:
    - df: DataFrame with synthetic prices
    """
    
    num_periods = len(market_data)

    # Step 1: Generate independent noise based on the specified distribution
    if distribution == 'normal':
        noise = np.random.normal(loc=0, scale=1, size=num_periods)  # Standard normal noise
    elif distribution == 'gamma':
        shape = 2  # Shape parameter for the gamma distribution
        noise = gamma.rvs(a=shape, scale=1, size=num_periods)
        noise = (noise - np.mean(noise)) / np.std(noise)  # Standardize noise
    elif distribution == 'lognormal':
        noise = lognorm.rvs(s=1, scale=np.exp(0), size=num_periods)
        noise = (noise - np.mean(noise)) / np.std(noise)  # Standardize noise
    elif distribution == 'uniform':
        noise = uniform.rvs(loc=-1, scale=2, size=num_periods)  # Uniform distribution [-1, 1]
    else:
        raise ValueError("Unsupported distribution. Choose from 'normal', 'gamma', 'lognormal', or 'uniform'.")

    # Step 2: Standardize market data
    market_std = (market_data - np.mean(market_data)) / np.std(market_data)

    # Step 3: Create a linear combination of market data and noise for correlated returns
    synthetic_log_returns = target_correlation * market_std + np.sqrt(1 - target_correlation**2) * noise

    # Step 4: Adjust the synthetic returns to have the desired mean and standard deviation
    synthetic_log_returns = synthetic_log_returns * sigma + mu

    # Step 5: Convert synthetic log returns to price data
    initial_price = np.random.uniform(5, 1000)  # Random initial price
    synthetic_prices = initial_price * np.exp(np.cumsum(synthetic_log_returns))

    # Step 6: Create a DataFrame with synthetic price data
    df = pd.DataFrame(synthetic_prices, columns=['Adj Close'])

    # Add High, Low, Volume columns to allow all functions to work
    df['High'] = df['Adj Close'] * (1 + np.random.normal(0, 0.01, size=len(df)))
    df['Low'] = df['Adj Close'] * (1 - np.random.normal(0, 0.01, size=len(df)))
    df['Volume'] = 0

    df = generate_technical_features(df)

    return df
\end{lstlisting}

The function generates synthetic log returns that are then converted to price data, while ensuring a target correlation with the real market data. This allows for a more controlled and realistic evaluation of the model's performance on different types of financial data.

\newpage

\section{Training Details}
The following sections give details on training procedures including both training of the models as well as optimal hyper parameters.
\subsection{qLSTM Hyperparamters}\label{app:qlstm-hyper}
The following table shows the optimal hyperparameters for the qLSTM model.

\begin{longtable}
{|p{5cm}|p{4cm}|p{4cm}|}
    \caption{Final Hyperparameters for the qLSTM model, rounded down to the 4th decimal place} \label{tab:final_lstm_hyperparameters}
        \\\hline
        \textbf{Hyperparameter} & \textbf{Asset Model (Stage 1)} & \textbf{Market Model (Stage 2)} \\ \hline
        \textbf{Batch size} & \multicolumn{2}{c|}{64} \\ \hline
        \textbf{Learning rate} & \multicolumn{2}{c|}{0.0006} \\ \hline
        \textbf{Normalization window} & \multicolumn{2}{c|}{219} \\ \hline
        \textbf{LSTM layers} & 1 & 1 \\ \hline
        \textbf{LSTM units} & 16 & 16 \\ \hline
        \textbf{Dense layers} & 5 & 3 \\ \hline
        \textbf{Dense units per layer} & 128, 64, 64, 32, 32 & 16, 16, 32 \\ \hline
        \textbf{Dropout rate} & \multicolumn{2}{c|}{0.1778} \\ \hline
        \textbf{Market activation} & \multicolumn{2}{c|}{Tanh} \\ \hline
        \textbf{Hidden activation} & \multicolumn{2}{c|}{ELU} \\ \hline
        \textbf{Use layer normalization} & \multicolumn{2}{c|}{no (0)} \\ \hline
        \textbf{L1 regularization} & \multicolumn{2}{c|}{0.0006} \\ \hline
        \textbf{L2 regularization} & \multicolumn{2}{c|}{0.0009} \\ \hline
\end{longtable}

\newpage

\subsection{qDense Training Details \& Results}\label{app:dense_training}

The following tables summarize the hyperparameters used during the training of the qDense model. The first table presents the ranges explored during the hyperparameter tuning process, and the second table lists the final hyperparameters used in the model, rounded down to the fourth decimal place.

\begin{longtable}
{|p{3cm}|p{4cm}|p{4cm}|}
    \caption{Hyperparameters explored for qDense Asset-Specific and Market Data Models}
    \label{tab:qdense_hyperparameters}
    
    \\ \hline
        \textbf{Hyper-parameter} & \textbf{Asset Model (Stage 1)} & \textbf{Market Model (Stage 2)} \\ \hline
        \textbf{Batch size} & \multicolumn{2}{c|}{\( b \in \{32, 64, 128, 256, 512, 1024, 2048\} \)} \\ \hline
        \textbf{Learning rate} & \multicolumn{2}{c|}{\( \eta \in [1 \times 10^{-6}, 1 \times 10^{-3}] \)} \\ \hline
        \textbf{Normalization window} & \multicolumn{2}{c|}{\( w \in [5, 250] \)} \\ \hline
        \textbf{Dropout rate} & \multicolumn{2}{c|}{\( D \in [0.0, 0.9] \)} \\ \hline
        \textbf{L1 regularization} & \multicolumn{2}{c|}{\( \lambda_1 \in [0.0, 1 \times 10^{-3}] \)} \\ \hline
        \textbf{L2 regularization} & \multicolumn{2}{c|}{\( \lambda_2 \in [0.0, 1 \times 10^{-3}] \)} \\ \hline
        \textbf{Use layer normalization} & \multicolumn{2}{c|}{Boolean: \( \{1, 0\} \)} \\ \hline
        \textbf{Raw hidden layers} & \( l_{\text{raw, asset}} \in [1, 5] \) & \( l_{\text{raw, market}} \in [1, 5] \) \\ \hline
        \textbf{Hidden layer units} & \( u_{\text{raw, asset}} \in \{16, \cdots, 512\} \) & \( u_{\text{raw, market}} \in \{16, \cdots, 512\} \) \\ \hline
        \textbf{Hidden activation function} & \(\{\text{relu}, \text{tanh}, \newline \text{sigmoid}, \text{leaky\_relu}, \text{elu}\}\) & \(\{\text{relu}, \text{tanh}, \newline \text{sigmoid}, \text{leaky\_relu}, \text{elu}\}\) \\ \hline
\end{longtable}

\begin{longtable}
    {|p{5cm}|p{4cm}|p{4cm}|}
    \caption{Final Hyperparameters for the qDense model, rounded down to the 4th decimal place}    \label{tab:final_qdense_hyperparameters}

        \\\hline
        \textbf{Hyperparameter} & \textbf{Asset Model (Stage 1)} & \textbf{Market Model (Stage 2)} \\ \hline
        \textbf{Batch size} & \multicolumn{2}{c|}{32} \\ \hline
        \textbf{Learning rate} & \multicolumn{2}{c|}{0.0003} \\ \hline
        \textbf{Normalization window} & \multicolumn{2}{c|}{207} \\ \hline
        \textbf{Raw hidden layers} & 3 & 1 \\ \hline
        \textbf{Hidden layer units} & 32, 32, 64 & 64 \\ \hline
        \textbf{Dropout rate} & \multicolumn{2}{c|}{0.4623} \\ \hline
        \textbf{Market activation} & \multicolumn{2}{c|}{ELU} \\ \hline
        \textbf{Hidden activation} & \multicolumn{2}{c|}{Leaky RELU} \\ \hline
        \textbf{Use layer normalization} & \multicolumn{2}{c|}{No (0)} \\ \hline
        \textbf{L1 regularization} & \multicolumn{2}{c|}{0.0005} \\ \hline
        \textbf{L2 regularization} & \multicolumn{2}{c|}{0.0001} \\ \hline
\end{longtable}

\end{document}